\documentclass[conference]{IEEEtran}

\IEEEoverridecommandlockouts

\usepackage{amsmath,amssymb,amsfonts}
\usepackage{algorithmic}
\usepackage[numbers]{natbib}
\usepackage{graphicx}
\usepackage{textcomp}
\usepackage{xcolor}
\def\BibTeX{{\rm B\kern-.05em{\sc i\kern-.025em b}\kern-.08em
    T\kern-.1667em\lower.7ex\hbox{E}\kern-.125emX}}
\begin{document}

\title{ Reducing Data Bottlenecks in Distributed, Heterogeneous Neural Networks\\
}

\author{\IEEEauthorblockN{Ruhai Lin}
\IEEEauthorblockA{\textit{Dept. of Electrical and Computer }\\ \textit{Engineering, UC Santa Cruz} \\
Santa Cruz, CA, United States \\
rlin50@ucsc.edu}
\and
\IEEEauthorblockN{Rui-Jie Zhu}
\IEEEauthorblockA{\textit{Dept. of Electrical and Computer }\\ \textit{Engineering, UC Santa Cruz} \\
Santa Cruz, CA, United States \\
rzhu48@ucsc.edu}
\and
\IEEEauthorblockN{Jason K. Eshraghian}
\IEEEauthorblockA{\textit{Dept. of Electrical and Computer }\\ \textit{Engineering, UC Santa Cruz} \\
Santa Cruz, CA, United States \\
jsn@ucsc.edu}

}

\maketitle

\begin{abstract}
The rapid advancement of embedded multicore and many-core systems has revolutionized computing, enabling the development of high-performance, energy-efficient solutions for a wide range of applications. As models scale up in size, data movement is increasingly the bottleneck to performance. This movement of data can exist between processor and memory, or between cores and chips. This paper investigates the impact of bottleneck size, in terms of inter-chip data traffic, on the performance of deep learning models in embedded multicore and many-core systems. We conduct a systematic analysis of the relationship between bottleneck size, computational resource utilization, and model accuracy. We apply a hardware-software co-design methodology where data bottlenecks are replaced with extremely narrow layers to reduce the amount of data traffic. In effect, time-multiplexing of signals is replaced by learnable embeddings that reduce the demands on chip IOs. 
Our experiments on the CIFAR100 dataset demonstrate that the classification accuracy generally decreases as the bottleneck ratio increases, with shallower models experiencing a more significant drop compared to deeper models. Hardware-side evaluation reveals that higher bottleneck ratios lead to substantial reductions in data transfer volume across the layers of the neural network. Through this research, we can determine the trade-off between data transfer volume and model performance, enabling the identification of a balanced point that achieves good performance while minimizing data transfer volume. This characteristic allows for the development of efficient models that are well-suited for resource-constrained environments. 
\end{abstract}


\section{Introduction}

Embedded multicore and many-core systems have fundamentally changed how we do high-performance computing. While neural networks are extremely good at exploiting many simple cores for highly parallel computation, the huge increase in the scale of models increases the volume of data that must be routed between these cores, potentially introducing bottlenecks for data movement. 

One key challenge in deploying deep learning models on embedded multicore and many-core systems is the efficient management and optimization of data transmission between layers. As the depth and complexity of models grow \cite{he2016deep, huang2017densely}, the limited memory bandwidth and computational resources of embedded systems can become significant bottlenecks, impacting both the speed and energy consumption of the system \cite{chen2017eyeriss, jouppi2017datacenter}.

In deep learning, the word `bottleneck' takes on a wholly distinct meaning. Bottleneck layers are commonly used to compress and encode information into a smaller feature maps, and are widely adopted in state-of-the-art models like ResNet \cite{he2016deep}, ResNeXt \cite{xie2017aggregated}, and MobileNetV2 \cite{sandler2018mobilenetv2}. By using encoding techniques such as dimensionality reduction bottleneck layers significantly decrease the amount of data transmitted between layers, reducing memory bandwidth requirements and enhancing computational efficiency \cite{sandler2018mobilenetv2, tan2019efficientnet}. As such, we propose to map algorithmic bottleneck layers to on-chip regions that can introduce hardware bottlenecks, such as IO pins. As such, bottleneck layers are a promising approach for addressing inter-chip and inter-core data movement in embedded multicore and many-core systems \cite{shao2020simba}. 

However, the optimization of bottleneck designs for such systems remains an open question, particularly in selecting the most suitable channel reduction ratio. Different bottleneck sizes, such as 32$\times$, 64$\times$, and 128$\times$, can have significant impacts on a model's performance and energy efficiency \cite{zagoruyko2016wide, tan2019efficientnet}. Finding the optimal balance between computational resource utilization and model accuracy is a key focus of this paper's exploration in the context of embedded systems.
Furthermore, considering the unique characteristics and constraints of embedded multicore and many-core systems, such as limited memory capacity and heterogeneous processing elements, the hardware adaptability of bottleneck layers is another crucial aspect of this research \cite{ma2018shufflenet, howard2019searching}. By investigating performance under different bottleneck size configurations, this paper aims to provide guidance to the embedded systems community on selecting the optimal bottleneck size based on different hardware architectures, enabling more efficient model deployment and execution on resource-constrained platforms \cite{shao2020simba, zhang2018dnnbuilder}.

The contributions of this paper can be summarized as follows:
\begin{enumerate}
    \item A systematic investigation of the impact of bottleneck size on the performance of neural networks, providing insights into the optimal balance between computational resource utilization and model accuracy.
    \item An analysis of the hardware adaptability of bottleneck technology, offering guidance on selecting the optimal bottleneck size based on different hardware architectures for efficient model training and inference processes.
    \item Leveraging insights from recent advancements in efficient neural network design \cite{howard2017mobilenets, iandola2016squeezenet}, this work provides guidance for the design of deep learning models.
\end{enumerate}

By addressing these key aspects, this paper aims to advance the understanding and application of bottleneck structures in deep learning, ultimately contributing to the development of more efficient and high-performing co-designed neural network models.

\section{Related Work}
\subsection{Bottleneck Architectures in Neural Networks}
Bottleneck architectures have been widely explored in the field of deep learning, particularly in the context of convolutional neural networks (CNNs). ResNet \cite{he2016deep}, one of the most influential CNN architectures, introduced the idea of residual connections and bottleneck layers to enable the training of deeper networks while mitigating the vanishing gradient problem. The bottleneck design in ResNet involves using 1$\times$1 convolutions to reduce and then restore the number of channels, which helps to reduce computational complexity and improve the efficiency of the network \cite{ zagoruyko2016wide}.

Building upon the success of ResNet, various architectures have incorporated and expanded on the bottleneck concept. ResNeXt \cite{xie2017aggregated} introduced the idea of aggregated transformations, which uses multiple parallel paths with bottleneck layers to increase the representational power of the network. DenseNet \cite{huang2017densely} employed dense connections between layers, where each layer receives the feature maps from all preceding layers, and uses bottleneck layers to control the number of input features. MobileNetV2 \cite{sandler2018mobilenetv2} and EfficientNet \cite{tan2019efficientnet} utilized inverted bottleneck structures, which expand and then reduce the number of channels, to build lightweight and efficient models for mobile and resource-constrained environments.

Recent works have further explored the optimization of bottleneck structures. ConvNeXt \cite{liu2022convnet} investigated the impact of various design choices in ResNet-like architectures and proposed a new architecture with inverted bottlenecks and depthwise convolutions. RepVGG \cite{ding2021repvgg} introduced a simple yet effective architecture that eliminates the need for explicit bottleneck structures by using a combination of 3$\times$3 and 1$\times$1 convolutions. These advancements demonstrate the ongoing efforts to optimize and simplify bottleneck designs for improved performance and efficiency.

\subsection{Hardware-aware Neural Network Design}
The co-design of neural network architectures and hardware has gained significant attention in recent years. Hardware-aware neural architecture search (HA-NAS) methods \cite{tan2019mnasnet, wu2019fbnet, cai2019once} have been proposed to automatically discover architectures that are optimized for specific hardware platforms, such as mobile devices or edge computing systems. These approaches consider hardware constraints, such as latency, memory usage, and energy consumption, during the architecture search process.

Most work in the co-design field focuses on the optimization of neural networks for specific hardware architectures, such as GPUs \cite{ma2018shufflenet, howard2019searching}, FPGAs \cite{zhang2018dnnbuilder, wang2019haq}, custom silicon-based accelerators \cite{chen2017eyeriss, jouppi2017datacenter, modaresi2023openspike}, and accelerators using emerging technologies~\cite{eshraghian2022memristor, wan2022compute}. These studies highlight the importance of considering hardware characteristics and constraints when designing and optimizing neural network architectures, including bottleneck structures.

In the context of bottleneck designs, hardware-aware optimization has been explored to improve the efficiency of neural networks on target hardware platforms. For example, Simba \cite{shao2020simba} proposed a multi-chip module (MCM) architecture that enables the design of processor core-specific bottlenecks to reduce inter-chip communication overhead. Other common approaches for reducing the strain of hardware bottlenecks aim to reduce data movement throughout the network, such as with sparsely activated models~\cite{he2024mixture} or spiking neural networks~\cite{eshraghian2023training, zhu2023spikegpt}.

\subsection{Efficient Neural Network Design}
Designing efficient neural networks has become increasingly important, particularly for deployment on resource-constrained devices and real-time applications. Strategies such as network pruning \cite{han2015learning, li2016pruning}, quantization \cite{hubara2017quantized, eshraghian2022navigating, jacob2018quantization, venkatesh2024squat}, low-cost operations~\cite{zhu2024scalable}, and knowledge distillation \cite{hinton2015distilling, romero2014fitnets, gunasekaran2024knowledge} have been proposed to reduce the computational cost and memory footprint of neural networks while maintaining acceptable performance.

In the context of bottleneck designs, various approaches have been explored to improve the efficiency of networks. MobileNetV1 \cite{howard2017mobilenets} and MobileNetV2 \cite{sandler2018mobilenetv2} introduced depthwise separable convolutions and inverted bottleneck structures, respectively, to build lightweight models for mobile devices. ShuffleNet \cite{zhang2018shufflenet} and ShuffleNetV2 \cite{ma2018shufflenet} employed channel shuffling and bottleneck structures to achieve high efficiency with reduced computational complexity. SqueezeNet \cite{iandola2016squeezenet} utilized fire modules, which consist of squeeze (reduction) and expand layers, to create compact networks with fewer parameters.
These works demonstrate the potential of incorporating efficient design principles, such as bottleneck structures, depthwise convolutions, and channel shuffling, to create compact and efficient neural networks. The development of hardware-friendly and resource-efficient architectures remains an active area of research, with implications for the deployment of deep learning models in real-world applications.

\section{Methods}
\subsection{Layers of Neural Network}
 In a neural network, each layer consists of multiple neurons (also called nodes). The data transmission between layers can be understood through the following steps:

\subsubsection{Input Layer}

The first layer of a neural network is the input layer, which receives the raw data. Suppose the input vector is $\mathbf{x} = [x_1, x_2, \ldots, x_n]$.

\subsubsection{Weights and Biases}

Each neuron is connected to all neurons in the previous layer through weights. Suppose the $l$-th layer has $n_l$ neurons, the weight matrix is $\mathbf{W}^{(l)}$, with dimensions $n_{l-1} \times n_l$ (the previous layer has $n_{l-1}$ neurons), and the bias vector is $\mathbf{b}^{(l)}$, with dimensions $n_l$.

\subsubsection{Linear Transformation}

Each neuron first performs a linear transformation, multiplying the input vector by the weight matrix and adding the bias. For the $j$-th neuron in the $l$-th layer, the linear transformation is represented as:
\begin{equation}
z_j^{(l)} = \sum_{i=1}^{n_{l-1}} W_{ij}^{(l)} a_i^{(l-1)} + b_j^{(l)}
\end{equation}
where $\mathbf{a}^{(l-1)}$ is the output (activation) of the $(l-1)$-th layer.
\subsubsection{Activation Function}
The result of the linear transformation is passed through an activation function (such as ReLU, Sigmoid, Tanh, etc.) to introduce non-linearity. Suppose the activation function is $f$, then the output of the $j$-th neuron is:
\begin{equation}
a_j^{(l)} = f(z_j^{(l)})
\end{equation}
\subsubsection{Output Layer}
This process repeats until reaching the output layer, whose output can be used for prediction or classification.
The entire process can be represented in matrix form as:
\begin{align}
\mathbf{Z}^{(l)} &= \mathbf{W}^{(l)} \mathbf{A}^{(l-1)} + \mathbf{b}^{(l)} \\
\mathbf{A}^{(l)} &= f(\mathbf{Z}^{(l)})
\end{align}
\subsection{Bottleneck}
\begin{figure*}
    \centering
    \includegraphics[width=0.7\textwidth]{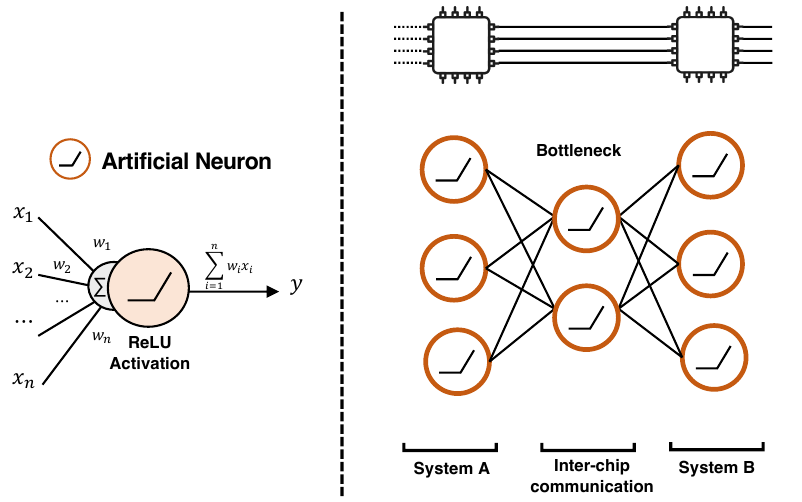}
    \caption{ Illustration of an artificial neuron with ReLU activation and the bottleneck structure in a multi-chip module (MCM) architecture. The bottleneck reduces inter-chip communication overhead between System A and System B.}
    \label{fig:main}
\end{figure*}

The bottleneck technique is a method to reduce the amount of data transmission and computation in neural networks. It is commonly used in CNNs in architectures like ResNet and Inception. The bottleneck structure is illustrated in Fig.~\ref{fig:main}.
\subsubsection*{Principle}
\begin{enumerate}
\item Dimensionality Reduction: Use a convolutional kernel (i.e., each kernel acts on only one pixel of the input feature map) to reduce the number of channels in the feature map. This reduces the computational load of subsequent convolutional layers.
\item Dimensionality Expansion: After reducing the dimensions, another convolutional kernel can be used to restore the number of channels in the feature map.
\end{enumerate}
Suppose the input feature map is $\mathbf{X}$ with dimensions $H \times W \times C_{in}$ (height, width, and number of channels). The feature map after dimensionality reduction is $\mathbf{Y}$ with dimensions $H \times W \times C_{mid}$. The process is as follows:
\subsubsection*{Dimensionality Reduction}
\begin{equation}
\mathbf{Y} = \mathbf{W} \mathbf{X} + \mathbf{b}
\end{equation}
where $\mathbf{W}$ is the convolutional kernel with dimensions $C_{in} \times C_{mid}$.
\subsubsection*{Dimensionality Expansion}
\begin{equation}
\mathbf{Z} = \mathbf{W}' \mathbf{Y} + \mathbf{b}'
\end{equation}
where $\mathbf{W}'$ is the convolutional kernel with dimensions $C_{mid} \times C_{out}$.
This method reduces the dimensions of the feature map, thereby lowering the computational complexity while maintaining the representational power of the features.

\subsection{Bottleneck in ResNet}
ResNet (Residual Network) is a popular CNN architecture that introduced the concept of residual connections and bottleneck structures. The bottleneck design in ResNet varies depending on the specific model:
\subsubsection*{BasicBlock (ResNet18, ResNet34)}
In ResNet18 and ResNet34, the BasicBlock consists of two convolutional layers with a residual connection. The bottleneck is applied between these two layers to reduce the number of channels. The structure of a BasicBlock with bottleneck is as follows:
\begin{enumerate}
\item 3$\times$3 convolution with $C_{in}$ input channels and $C_{mid}$ output channels;
\item Batch Normalization and ReLU activation
\item 3$\times$3 convolution with $C_{mid}$ input channels and $C_{out}$ output channels;
\item Addition of the residual connection (if dimensions match);
\item Batch Normalization and ReLU activation.
\end{enumerate}
\subsubsection*{Bottleneck Block (ResNet50, ResNet101, ResNet152)}
In deeper ResNet models (ResNet50, ResNet101, and ResNet152), the Bottleneck Block consists of three convolutional layers with a residual connection. The bottleneck is applied in the middle layer to reduce the number of channels. The structure of a Bottleneck Block is as follows:
\begin{enumerate}
\item 1$\times$1 convolution with $C_{in}$ input channels and $C_{mid}$ output channels;
\item Batch Normalization and ReLU activation;
\item 3$\times$3 convolution with $C_{mid}$ input channels and $C_{mid}$ output channels (bottleneck layer);
\item Batch Normalization and ReLU activation;
\item 1$\times$1 convolution with $C_{mid}$ input channels and $C_{out}$ output channels;
\item Addition of the residual connection (if dimensions match);
\item Batch Normalization and ReLU activation.
\end{enumerate}
The bottleneck ratio, which determines the reduction in the number of channels, is applied in the middle layer of the Bottleneck Block. By reducing the number of channels in this layer, the computational complexity is decreased while still allowing for the learning of rich features.
The use of bottleneck structures in ResNet has proven to be highly effective in achieving state-of-the-art performance on various computer vision tasks. By carefully designing the bottleneck ratio and the overall architecture, ResNet models can strike a balance between model complexity and representational power, enabling them to learn deep and informative features from the input data.

\section{Experiments}
We conduct performance tests to evaluate the impact of bottleneck size on the classification accuracy and hardware efficiency of ResNet models. The experiments are divided into two parts: software-side evaluation using the CIFAR100 dataset and hardware-side evaluation focusing on communication latency.
\subsection{Software-side Evaluation}
\subsubsection{Dataset}
CIFAR100 is a widely used benchmark dataset in the field of computer vision and machine learning. It consists of 60,000 color images with a size of 32x32 pixels, divided into 100 classes, with 600 images per class. The dataset is split into 50,000 training images and 10,000 test images, providing a challenging task for image classification algorithms due to the high number of classes and the limited number of samples per class \cite{krizhevsky2009learning}.
The CIFAR100 dataset is an extension of the CIFAR10 dataset, which contains 10 classes with 6,000 images per class. CIFAR100 introduces a hierarchical labeling scheme, where each class belongs to one of 20 superclasses. This hierarchical structure allows for the evaluation of models at different levels of granularity and enables the exploration of techniques such as hierarchical classification and transfer learning \cite{zeiler2014visualizing, oquab2014learning}.

The images in CIFAR100 cover a wide range of object categories, including animals (e.g., mammals, fish, insects), vehicles (e.g., cars, trucks, planes), and everyday objects (e.g., furniture, fruits, household items). The dataset's diversity and complexity make it an ideal testbed for evaluating the performance of deep learning models, particularly CNNs \cite{he2016deep, huang2017densely}.

In our experiments, we use the CIFAR100 dataset to assess the impact of bottleneck size on the classification accuracy of ResNet models. By training and evaluating ResNet architectures with different bottleneck configurations, we aim to provide insights into the optimal design choices for achieving high performance on this challenging dataset.

\subsubsection{Experimental Setup}
We investigate the impact of bottleneck ratio on the classification accuracy of various ResNet models, including ResNet18, ResNet34, ResNet50, ResNet101, and ResNet152. The bottleneck ratios considered in our experiments are 1, 2, 4, 8, 16, and 32. These values represent the ratio by which the number of channels is reduced in the bottleneck layers. A higher bottleneck ratio corresponds to a smaller number of channels after the squeezing operation.
All models are trained and evaluated on the CIFAR100 dataset using the same training hyperparameters and data augmentation techniques to ensure a fair comparison. The models are trained for a fixed number of epochs, and the classification accuracy on the test set is reported.
\subsubsection{Results and Analysis}
\begin{figure}
    \centering
    \includegraphics[width=\linewidth]{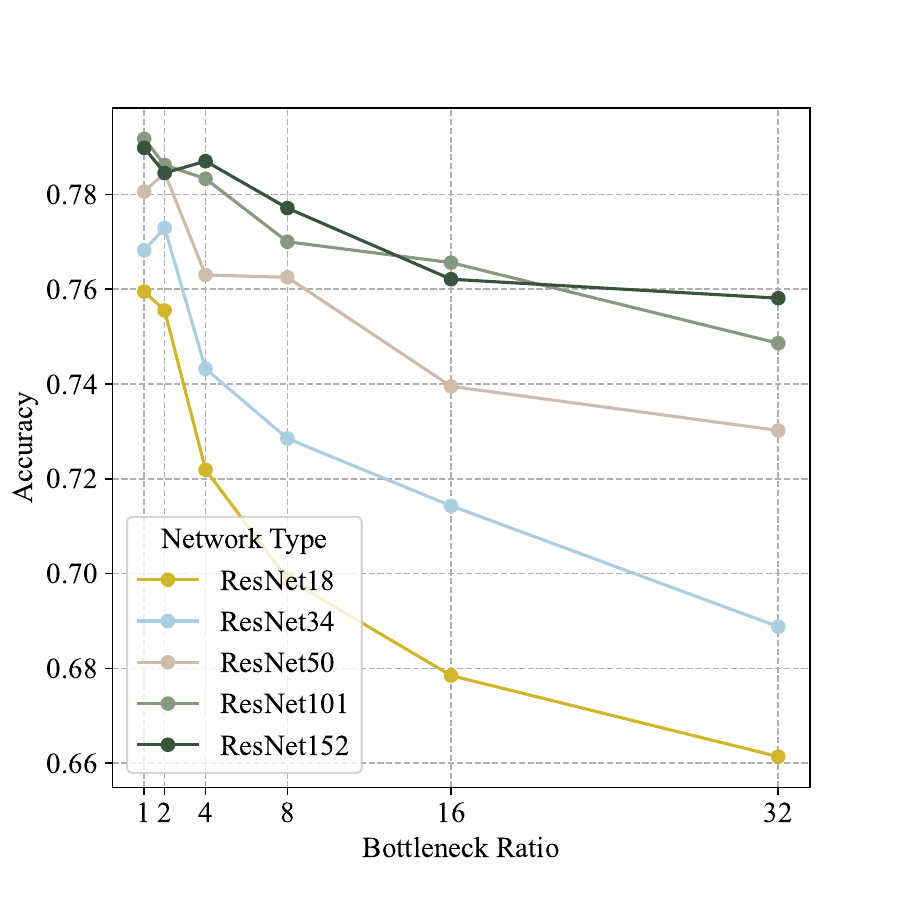}
    \caption{Accuracy vs Bottleneck for Different ResNet Architectures.}
    \label{fig:acc}
\end{figure}
From the results, we observe that the classification accuracy generally decreases as the bottleneck ratio increases for all ResNet models. This trend suggests that higher bottleneck ratios, which correspond to more aggressive channel reduction (smaller bottleneck sizes), may lead to a loss of information and a decrease in the model's representational capacity.

However, the impact of bottleneck ratio varies across different ResNet architectures. Shallower models, such as ResNet18 and ResNet34, experience a more significant drop in accuracy as the bottleneck ratio increases compared to deeper models like ResNet101 and ResNet152. This observation indicates that deeper models are more resilient to the information loss caused by higher bottleneck ratios, likely due to their increased depth and ability to capture more complex features.

Interestingly, for ResNet34 and ResNet50, the highest accuracy is achieved with a bottleneck ratio of 2, suggesting that a moderate level of channel reduction is beneficial for performance. This finding aligns with the concept of the inverted bottleneck structure used in architectures like MobileNetV2 \cite{sandler2018mobilenetv2}, which employs expansion and reduction factors to balance computational efficiency and representational power.

\subsection{Hardware-side Evaluation}

\begin{figure}
    \centering
    \includegraphics[width=\linewidth]{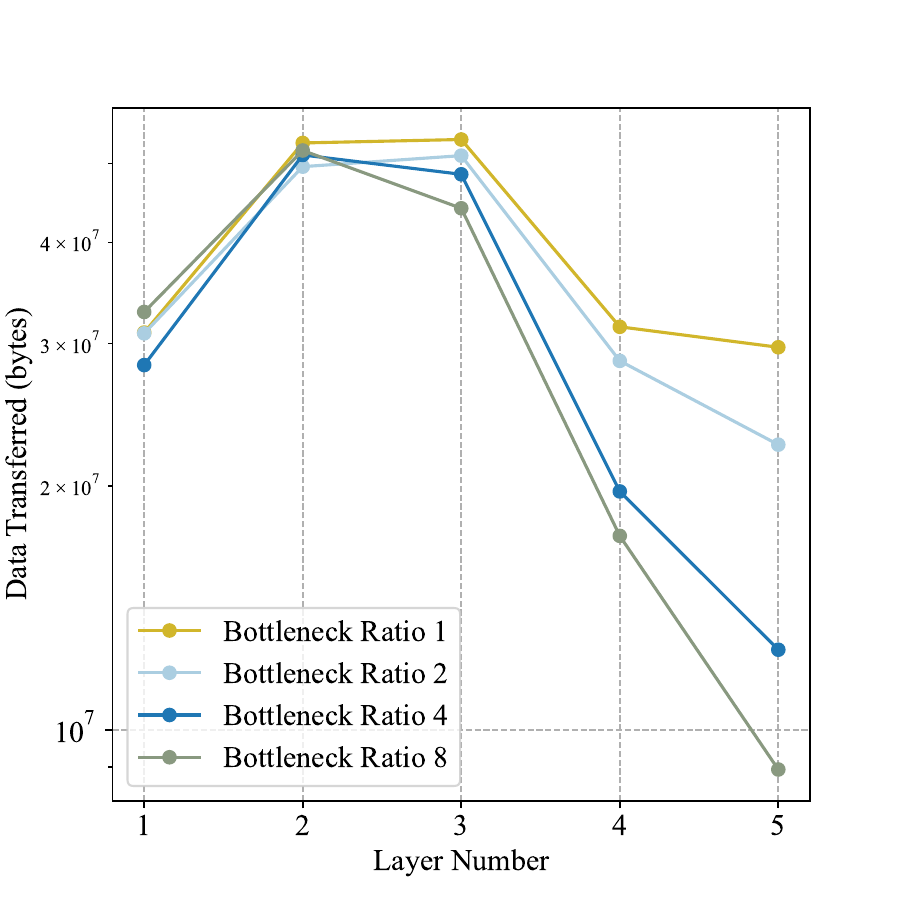}
    \caption{Data Transferred vs Layer Number for Different Bottleneck Ratio.}
    \label{fig:acc}
\end{figure}

\subsubsection{Experiment Setup}
On the hardware verification side, we used the ResNet-18 model in our experiment. The initial parameters were chosen to be straightforward and random. We used PyTorch's torch.randn function to generate a random tensor with the shape (1, 3, 224, 224), simulating an input batch containing a single 224x224 RGB image.

To evaluate the impact of the bottleneck technique, each layer of the neural network was assigned to a separate thread. This ensured that the computations were carried out linearly—each layer's output was fed into the subsequent layer only after its computation was completed. During this process, we monitored the memory usage of each thread using the virtual\_memory().used function from the psutil library. When one layer of the neural network is about to start the computation, the program will record the total memory usage once, and when the layer completes the calculation, the program will record the memory usage again and calculate the difference between the two measurement values, which gives the data transmission volume of this layer.

\subsubsection{Result and Analysis}
The effectiveness of the bottleneck method was demonstrated by the reduction in the amount of data transferred of each layer, i.e., the memory usage. The results indicated significant reductions in data transfer across different bottleneck ratios. As the number of layers increases, the advantages brought by bottleneck technology and the reduced amount of data transmission become more and more obvious. Compared to Bottleneck Ratio 1, Bottleneck Ratio 8 reduced the data transfer amount by approximately 70\% at the fifth layer already. The data transfer per layer for different bottleneck ratios is depicted in Figure 2, showcasing the memory usage trends across the layers of the ResNet-18 model.

\section{Conclusions}
In this work, we systematically investigated the impact of bottleneck size on the performance and efficiency of neural network models in embedded multicore and many-core systems. Through software-side experiments on the CIFAR100 dataset, we observed that classification accuracy generally decreases as the bottleneck ratio increases, with shallower models experiencing a more significant drop compared to deeper models. Hardware-side evaluation revealed that higher bottleneck ratios lead to substantial reductions in data transfer volume across the layers of the ResNet-18 model. These findings enable us to determine the trade-off between data transfer volume and model performance, allowing for the development of efficient models well-suited for resource-constrained environments. The insights gained from this study contribute to the ongoing efforts in designing efficient neural network architectures for embedded systems and provide valuable guidance for selecting the optimal bottleneck size based on specific hardware architectures and performance requirements. By leveraging these findings, researchers and practitioners can develop high-performing, energy-efficient solutions that push the boundaries of AI capabilities on resource-constrained platforms, paving the way for the widespread deployment of deep learning models in embedded systems.

\section*{Acknowledgement}
This project was supported in-part by the National Science Foundation RCN-SC 2332166.

\bibliographystyle{IEEEtran}

\vspace{12pt}

\end{document}